\documentclass[10pt,a4paper]{article}
\title{\bf SOCIAL BEHAVIOR OF DRIVERS \\ AND TRAFFIC.}
\author{Matej Hudak \\ {\it Lab,} Stierova 23, SK - 040 23  Kosice, Slovak Republic  \\hudakm@mail.pvt.sk
	\and Jana Tothova\\ {\it Lab,} Stierova 23, SK - 040 23  Kosice, Slovak Republic
	\and Ondrej Hudak \\  {\it Technical University of Kosice},\\ Faculty of Aerodynamics, Department of Aviation Technical Studies,\\ Rampova 7, SK - 040 01  Kosice, Slovak Republic\footnote{Corresponding author}\\
	{hudako@mail.pvt.sk}}
\date{April 16, 2018}

\begin{document}
	\maketitle{}

	\section*{PACS Numbers:}
	\begin{itemize}
		\item 
		89.75.-k 	Complex systems
		\item 
		89.70.Cf 	Entropy and other measures of information
		\item 
     	89.65.Lm 	Urban planning and construction 
		\item 
		7.85.G- 	Biomechanics
		\item 
		91.10.Ws 	Reference systems (also on the planets and asteroids - for motion)
	\end{itemize}
	\section*{Condensed paper title:}
	{\it BEHAVIOR OF DRIVERS AND TRAFFIC.}
	\newpage
	\vspace{1in}
	\begin{abstract}
	Drivers are agents, they are members of the group of drivers. Human groupings have hierarchical structure. The civilization consists of societies, societies consist of groups, and groups consist of individuals. We will consider the group of drivers. Every social group has power to organize individuals and use them for its own purposes. This is true also for the group of drivers. An agent, a driver, we call a member of a social group of drivers, we assume that his-her special properties are defined: they are agents driving a car.
	A driver´s behavior is influenced to some degree by the need to associate with other drivers and to obtain the approval of other drivers in the group, this is a general property of agents in a group.  A driver equates his-her needs with those of the other drivers from the group.
	From this we explain how three empirically observed dependencies
	of personal driver radius dependence on some factors enabled us to identify quantity which characterizes verification of information by a driver. The smaller personal radius the larger process of verification of verbal and non-verbal information about the other person in other to accept the risk that this person will be closer to us as a person (driver). We expect that this conclusion is a general conclusion.

	\end{abstract}
	\newpage
	
	\tableofcontents{}

\newpage
\section{Introduction}
We have recently studied \cite{OH} connection between topology and social behavior of agents (O. Hudak). This study was applied to capital markets \cite{OH1}. In \cite{JOH} and \cite{JOH1}partially this method was used in study of     firm projects, NPV and risk. And in \cite{OH2} we studied social behavior of agents on capital markets and their small perturbations (O. Hudak). We will use this characterization of a group of agents to drivers. Let us in the following describe the method mentioned above and then apply it to drivers and their social behavior 

We call in a social group its members as agents. 
In general the social group is characterized in accordance with 
T. Plummer \cite{1} . We assume that individual behavior 
of agents is influenced to some degree by the need to associate 
with other agents and to obtain the approval of other agents in 
the group. The group is characterized by a very large non-rational 
and emotional element to decisions of agents. It is due to the fact
 that making decisions an individual equates own needs with those
 of the other agents from the group. Any two agents from the group 
may interact. The interaction consists of the exchange of information 
and it costs some energy. The information is well defined, and we 
assume that agents interact in such a way that they give reference 
to the origin of the information if asked by other agents. Thus 
the agent may verify obtained  information. It is natural then 
that there exits a subgroup of interacting agents the interaction 
of which has the following property: it is non-reducible in the sense 
that in this subgroup there exists the interaction of a given agent 
with another one, this another one agent again interacts with another
 agent, and because it has no sense to exchange the information with 
the first one to verify the information, this last mentioned agent is 
different from the first one. This third agent can thus always verify 
the information from two sources either interacting with the first one,
 either he-she is interacting with the fourth agent. In the first case
 we have a minimal subgroup of interacting agents, which form a closed 
subgroup in which every agent verifies information from two
sources - agents. In the second case the fourth agent either interacts
 with the first one and verifies exchanged information, either with 
the second one and verifies information. He of course can interact 
with the third one agent, however to verify information he needs 
again to interact with different agent from that interacting with 
which he obtained the new information. This is the reason why he 
interacts either with the first one agent, either with the second one. 
In the first case we have a closed subgroup of four agents in which 
every agent interacts with two agents. In the second case the closed
 subgroup reduces to the group of  three interacting agents (last 
three agents). This process can continue further, because the fourth 
agent may also verify information interacting with the fifth agent. 
Then by the same procedure as in previous case we may obtain 
non-reducible subgroups of three, of four and of five agents, or 
the fifth agent interacts with the sixth one. And so on. Thus we 
obtain closed subgroups of interacting agents in which every agent
 interacts with two other agents, and in which  the process of 
verification of information leads to closed linear structure. If any
 of agents from such an subgroup exchange and thus verifies information
 with any third one agent from the subgroup, the structure of 
the subgroup changes, the subgroup becomes reducible to two
 new non-reducible subgroups. This is the process of differentiation 
of the first non-reducible subgroup to two new ones. This is an 
example of an elementary transformation between two configurations 
of non-reducible subgroups An vice versa: if the interaction between 
two agents which interact with three agents in the configuration of 
two non-reducible subgroups vanishes, a new one non-reducible subgroup
 creates and information is still verifiable. When the configuration 
in which an agent was interacting with three agents which were not
 interacting between themselves transform to a configuration in which 
two of the mentioned three agents interact, then we may observe 
the process of mitosis: a new non-reducible group appears.Thus 
the transformation between this two configurations is reversible 
A cell is such a configuration of a given number of non-reducible 
subgroups in which every two interacting agents belongs to two 
non-reducible subgroups (subgroups are connected in this sense)
and which is closed . Such a cell  may disappear and may be created,
may change number of non-reducible subgroups in a reversible way.
 Because the structure, configuration of interactions between  
agents in the group, forms a macroscopic structure, we say that it is 
a micro-reversible process any process within a non-reducible subgroup 
and within a cell. Statistical equilibrium of the whole group 
is characterized by a set of different subgroups of the type mentioned
 above, and by a probability that such a subgroup occurs. Thus we have
 probability distribution which characterizes the group. Moreover 
there exists an equation of state which enables to compare different
 macroscopic states of the group. The statistical equilibrium due
 to micro-reversibility is characterized by the maximum of entropy 
and by the minimum of energy (costs of information exchange). 
We will use methods of statistical physics to study  social behavior of agents, mainly the presence of topological structure of interactions between 
agents and its changes, which is the most important property of
 the group of agents. There are three empirically observed dependencies of personal radius which enabled us to characterize the quantities of cells, faces, vertices and bonds \cite{2}.
There exist constrains, such as a fixed number V of agents in 
the group,  a number E of interactions within the group, a number F 
of subgroups which are non-reducible, and a number C of cells. Thus 
we have a structure which is equivalent to random cellular networks. 
Such networks and their evolution were described by N. Rivier
 \cite{3} and \cite{4}. He applied methods of statistical mechanics to study these structures. We will use methods described by Rivier to study  social behavior of agents, mainly the presence of topological structure of interactions between agents and its changes, which is the most important property of the group of agents. The area of a non-reducible group which belongs to those non-reducible groups which form the cell may be formed again for example by a sum of areas of  agents characteristic areas. Note 
that area of the non-reducible group may be also some other 
characteristics of the group of agents depending on studied social 
relations between agents. Thus we are able to study topology 
properties of interactions of agents. Their social behavior is
 discussed in. It can be shown that the equilibrium number of agents 
with which a given agent interacts is three for a group without 
cells (the group forms a single cell).

\section{Human groupings have hierarchical structure.} 

Human groupings have hierarchical structure, for an introduction see
\cite{1}. The civilization consists of societies, societies consist of groups, and groups consist of individuals. A social group has power to organize individuals and se them for its own purposes.
An agent we call a member of a social group, we assume that his-her special properties are defined.

Agent - its behavior is influenced to some degree by the need to associate with other agents and to obtain the approval of other agents
in the group \cite{1}.
A group is characterized by a very large non-rational and emotional
elements in decisions of agents, and agent equates his-her 
needs with those of the other agents from the group [1]. 
Equates here means interacts, interaction costs some energy and consists of the exchange of information about his-her needs, the information is well defined. Specification of agents here: agents interact in such a way that they give reference to the origin of the information if asked by the other agent. Thus an agent may verify obtained information.

Non-reducible subgroup of the group of agents is a group which defined in the following way.
There exists interaction of a given agent with another one. This
another one agent interacts with another one. it has no sense for him-her to exchange information with the first one to verify
information (to obtain approval of at least another one agent):
thus the agent interacts with and agent different from the first one.
This the third agent can verify the information from two sources: 
either interacting with the first one, either interacting with the fourth one.
The first case leads to formation of a non-reducible subgroup of three 
agents in which every agent verifies information from two different 
sources (we assume that information is verified if an agent verifies
information from at least two different sources).
The second case: the fourth agent interacts either with the first
one, and we obtain a non-reducible subgroup of four agents, either this agent interacts with the second agent,
and we obtain a non-reducible subgroup of three agents, this process
may continue further: the fourth agent may verify information
interacting with the first ones, the process just describes now may
lead to a non-reducible subgroup with three, four or five agents,
etc. Thus in the group of interacting agents there exist non-reducible
subgroups of agents which are closed as concerning exchange of an information and in which every agent interacts with two
and only two other agents, thus the structure of interaction of agents is "linear-circular".

\section{Non-reducible subgroups of the group of agents may transform.}

Reduction of a non-reducible subgroup to two non-reducible subgroups:
if any agent from a given non-reducible subgroup interacts as
 concerning exchange of an information 
with a third one agent from the non-reducible subgroup.
And vice versa: if the interaction as concerning exchange of an information between two agents in the configuration of two non-reducible subgroups vanishes, a new one non-reducible subgroup appears. Mitosis is a process in which an agent which was interacting with three agents which were not interacting between themselves transforms to a configuration in which two of the mentioned three agents interact, then a new non-reducible subgroup appears, this process is reversible

Cell is a configuration of a given number of non-reducible subgroups
in which every two interacting agents belong to two non-reducible
subgroups of a closed subgroup formed from non-reducible subgroups.
Cell of the group of agents may reversible transform, they may 
disappear, may be created, may change a number of non-reducible
subgroups.

\section{The structure of the group is macroscopic}

The group has a given number of agents, of non-reducible subgroups and of cells as concerning exchange of information. Statistical 
equilibrium of the group is characterized by different macroscopic 
structure as concerning exchange of an information and by a probability that a non-reducible group appears, this probability characterizes the group. Equation of state comparison of different 
macroscopic states of the group. Statistical equilibrium exist in the group due to reversibility, it is characterized by the maximum of entropy and by the minimum of energy (costs of information exchange).
Constraints are given by a fixed number V of agents in the group, a number E of interactions, a number F of subgroups which are non-reducible, a number C of cells, note that conservation law holds \cite{4}:
\begin{equation}
\label{1}
- C + F - E + V = 0
\end{equation}

Note that structural stability exists \cite{4}: only agents with interaction with four agents are structurally stable in 3d, and only agents with interaction with three agents are structurally stable in 2d. This structure is equivalent to random cellular networks,
such networks and their evolution were described by N. Rivier \cite{3}, \cite{4} by methods statistical mechanics.

We will use these methods to study social behavior of agents,
and changes of this behavior, mainly the presence of the topological
structure of interactions between agents and its changes, the most 
important property of the group of agents which corresponds to maximum 
of informational entropy.

One can define an area of non-reducible subgroup and volume of the cell. We assume that there is homogeneity, and no costs of information.

\section{Personal area and social behavior of agents of the group}

The average area A(n) of an n-sided cell is \cite{3}, \cite{4} (if A is the total area in which group is localized):
\begin{equation}
\label{2}
A(n) = \frac{A}{F}\lambda (n-(6-\frac{1}{\lambda}))
\end{equation}

Assume that this area corresponds to a personal area of agents forming 
an n-sided non-reducible subgroup, due to homogeneity and equilibrium
every agent contributes $\frac{1}{3}$ personal area A(n), thus $ \lambda = \frac{1}{6}$ and
\begin{equation}
\label{3}
A(n) = \frac{A}{F} \frac{n}{6}
\end{equation}

One agent contributes $ \frac{1}{3} $ of his personal area to the area
A(n), if we denote r the radius of the agent's personal area then:
\begin{equation}
\label{4}
\frac{\pi r^{2}}{3} = \frac{A}{6F}
\end{equation}

As we see personal radius increases as a square root with increased

total area A per a non-reducible group. As we see personal radius
decreases as an inverse square root with increasing number of
non-reducible groups, in towns personal area is observed to be smaller
than in villages \cite{2}, this corresponds to smaller $ \frac{A}{F} $
ratio in towns than in villages according to our equation above,
the ratio $ \frac{A}{F} $ is more-less constant in area A in these cases, however it is increasing with the density of agents in the group (town, village) and thus number of non-reducible groups is increasing in our equation for personal area qualitatively, personal radius is increasing with decreasing risk which a person expects
\cite{2}: intimous - smaller radius is 0.15 m and less for intimous contacts (love, security, ... ), intimous - larger radius is 0.15 m to 0.45 m for less intimous contacts (relatives), personal - smaller radius is 0.45 m to 0.90 m for personal contacts with close friends and relatives, personal - larger radius is 0.90 m to 1.20 m for personal contacts with friend, business people, neighbors
Thus we see from our equation that F may be associated with risk:
for constant area A the larger F the smaller r and smaller acceptable risk (*). Personal radius is large for inhabitants \cite{2} of New Zealand, Australia and white North-Americans, it is middle for inhabitants of Great Britain, Switzerland, Sverige, Germany, Austria
and it is small for inhabitants of Arab countries, Japans, South-Americans, inhabitants of countries around Middleterrenian sea (Italy, France, Greece, ... ) and black North-Americans: one can say qualitatively that (probably due to temperament) the first group is characterized by low risk activities and by their preference, the second group is characterized by middle risk activities, and the last group is characterized by high risk activities. Thus first group has F lower than the second group, and second group lower than the third group: this is consistent with the statement (*). 

\section{Social Behavior of drivers}
Drivers are agents, they are members of the group of drivers. Human groupings have hierarchical structure, \cite{1}. The civilization consists of societies, societies consist of groups, and groups consist of individuals. We will consider the group of drivers. Every social group has power to organize individuals and use them for its own purposes. This is true also for the group of drivers. An agent, a driver, we call a member of a social group of drivers, we assume that his-her special properties are defined: they are agents driving a car.

A driver´s behavior is influenced to some degree by the need to associate with other drivers and to obtain the approval of other drivers in the group, this is a general property of agents in a group \cite{1}. The need to associate with other drivers follows from the fact that they are driving on the same streets and ways, or highways, they should preserve the same rules for driving, etc. A group of drivers is characterized by a very large non-rational and emotional elements in decisions of drivers, as in other groups, and a driver equates his-her needs with those of the other drivers from the group \cite{1}. Equates here means interacts, interaction consists of the exchange of verbal and nonverbal  information about his-her needs, the information which drivers exchange is well defined, we assume here in this paper. In fact this may be not true. Agents interact in such a way that they give reference to the origin of the information if asked by the other agent, for example at the law trial. In fact a driver has to take into account the signals of a driver in a car in front of him, behind him, on both sides of the car, and signals of traffic lights (as a first approximation).

Thus an agent may verify obtained information only in a line of agents. Thus drivers in cars are forming lines, these lines are starting and ending at the points where the street, the way or the highway is splitting. This point is the point called a vertex in topology.

Non-reducible subgroup of the group of agents in lines (in streets, in ways, in highways) is a group which is defined in the following way.
There exists interaction of a given line of drivers with another lines.
This another one line of drivers interacts with another one. It has no sense for the drivers in a given line to exchange information with the first line of drivers to verify
information (to obtain approval of at least another one line of drivers).
The lines of drivers interact with another line of drivers different from the first one. The third line of drivers agent can verify the information from the two sources: either interacting with the first line of drivers, either interacting with the fourth one. The first case leads to formation of a non-reducible subgroup of three lines of drivers in which every line verifies information from two different sources (we assume that information is verified if an line of drivers verifies information from at least two different sources - lines of drivers). The second case: the fourth line of drivers interacts either with the first one, and we obtain a non-reducible subgroup of four lines of drivers, either this line of drivers interacts with the second line of drivers, and we obtain a non-reducible subgroup of three lines of drivers, this process may continue further: the fourth line of drivers may verify information interacting with the first one, the process just describes now may lead to a non-reducible subgroup with three, four or five lines of drivers, etc. Thus in the group of interacting lines of drivers there exist non-reducible subgroups of lines of drivers which are closed as concerning exchange of information and in which every line of drivers interacts with two and only two other lines, thus the structure of interaction of agents is "linear-circular".

Reduction of a non-reducible subgroup of lines of drivers to two non-reducible subgroups of lines of drivers is the following process: if any line of drivers from a given non-reducible subgroup interacts as concerning exchange of an information with a third one line of drivers from the non-reducible subgroup (for example opening a street, a way or a highway, which is joining these two lines at vertices by changing the traffic signals, or by other way). And vice versa: if the interaction as concerning exchange of an information between two lines of drivers in the configuration of two non-reducible subgroups vanishes, and  a new one non-reducible subgroup appears when a street, a way or a highway is closing by traffic signals or for other reasons. Mitosis is a process in which an line of drivers which was interacting with other two lines of drivers, which were not interacting between themselves, transforms to a configuration in which two of the mentioned three lines of drivers start to interact, then a non-reducible subgroup disappears, this process is reversible

\section{Conclusion}

We conclude that the three mentioned empirically observed dependencies
of personal radius dependence on some factors enabled us to characterize the quantity F as the quantity which characterizes verification of information, and the smaller personal radius the larger process of verification of verbal and non-verbal information about the other person in other to accept the risk that this person will be closer to us as a person. We expect that this conclusion is a general conclusion. One may expect that on systems (planets, asteroids) with vehicles in motion (planes on Earth f.e., vehicles in future on Luna or Mars or asteroids) the reference systems (of 2d or of 3d type) should exists ("highways") on which automatic or men drivers will drive these vehicles. Then our description above may be generalized to these systems.

\end{document}